\documentclass[pra,twocolumn,showpacs,preprintnumbers,amsmath,amssymb]{revtex4}
\usepackage{mathrsfs}
\usepackage{hyperref}
\def\sp#1{\mathscr #1}
\def\N#1{|\!|#1|\!|}
\def\kk{\rangle\!\rangle}\def\bb{\langle\!\langle}
\def\d{\operatorname{d}}
\def\transp#1{{#1}^\tau}
\def\SU#1{\mathbb{SU}(#1)}
\def\sH{\sp H}
\def\set#1{\mathscr #1}
\def\out{\set X}

\def\Cmplx{\mathbb C}
\def\<{\langle}\def\>{\rangle}
\def\Tr{\mbox{Tr}}
\def\bP{{\bf P}}\def\bE{{\bf E}}\def\bQ{{\bf Q}}\def\bF{{\bf F}}
\def\Pn{\mathscr{P}_n}\def\PF{\mathscr{P}_{\bf F}}\def\anc{\mathscr{A}}
\def\map#1{\mathcal{#1}}\def\dim{\operatorname{dim}}
\begin{document}
\title{Efficient universal programmable quantum measurements}
\author{Giacomo Mauro D'Ariano}\email{dariano@unipv.it}
\altaffiliation[Also at ]{Center for Photonic Communication and Computing, Department of
  Electrical and Computer Engineering, Northwestern University,
  Evanston, IL 60208}
\author{Paolo Perinotti}\email{perinotti@fisicavolta.unipv.it} 
\affiliation{{\em QUIT Group}\homepage{http://www.qubit.it}, Dipartimento di Fisica
  ``A. Volta'', via Bassi 6, I-27100 Pavia, Italy} \date{\today}
\begin{abstract}
A universal programmable detector is a device that can be tuned to perform any desired measurement
on a given quantum system, by changing the state of an ancilla. With a finite dimension $d$ for the ancilla
only approximate universal programmability is possible, with ``size'' $d=f(\varepsilon^{-1})$ increasing
function of the ``accuracy'' $\varepsilon^{-1}$. In this letter we show that, much better than the exponential size 
known in the literature, one can achieve polynomial size. An explicit example with linear size is
also presented. Finally, we show that for covariant measurements exact programmability is feasible. 
\end{abstract}
\maketitle
A concrete problem in quantum information processing \cite{Nielsen} is to experimentally achieve any
theoretically designed quantum measurement, and possibly be able to change the measured observable
dynamically on the fly, as it would be needed e.~g. when trying to eavesdrop quantum-encrypted
information.  For such a purpose, a programmable measurement apparatus, which could be tuned to
perform any desired measurement, would be an invaluable resource. However, as first noticed in
Refs. \cite{buzek,fiurasek1}, with a finite dimensional ancilla, exact universal programmability of
measurement is impossible, as a consequence of the no-go theorem for programmability of unitary
transformations \cite{niels}. One can  still achieve measurement programmability probabilistically,
or even deterministically, though within some accuracy. Since different measurements within some
classes can be mapped to each other via quantum channels (e.~ g. all observables are connected
each other by a unitary transformation), then the problem of measurement programmability clearly
carries relations with that of channels programming\cite{ciravid}. Thanks to the correspondence
between channels and bipartite states\cite{choi,clon_cov}, quantum channels can be easily
programmed probabilistically by using a teleportation scheme with the channel stored in the state of
the shared bipartite resource\cite{Nielsen}.  This and other methods can then be used to program
channels and measurements probabilistically\cite{buzekproc,fiurasek2}, and recently optical
implementations for polarization-encoded qubits have been proposed\cite{fiurasek3}. However, one
should emphasize that, differently from the case of programmability of quantum channels or
operations---where a series of many of them in sequence amplifies errors---in the  case of a quantum
measurement, being the last quantum processing stage it is certainly more efficient to consider
deterministic programmability at the expense of small bounded systematic errors, rather than
achieving the exact measurement probabilistically.  In Ref. \cite{fiurasek1} a measurement for
qubits that can be approximately programmed to achieve any observable has been presented, 
which needs an ancilla with dimension growing exponentially versus the accuracy $\varepsilon^{-1}$. 
In this letter we will show that actually it is possible to design the programmable measurement
much more efficiently, with dimension $d$ of the ancilla growing only polynomially versus the accuracy
$\varepsilon^{-1}$.   We will also provide a specific example for such efficient programmability for
qubits, with dimension $d$ linear in $\varepsilon^{-1}$. We also show that in some cases, e.~g. when
the programmability is restricted to covariant measurements, even exact deterministic
programmability is possible.   

In quantum mechanics the statistics of a generic measurement apparatus is described by a positive
operator-valued measure (POVM). For simplicity in this letter we will consider the case of discrete
sampling space $\out$ of possible outcomes for the measurement, in which case a POVM  $\bP$ is just
a set of positive operators $P_i\geqslant 0$ on the Hilbert space $\sH$ of the system, each
corresponding to an elementary outcome $i\in\out$, and satisfying the normalization condition
$\sum_{i\in\out}P_i=I$. In the following to simplify notation, we will write simply $\sum_i$ for
$\sum_{i\in\out}$, and don't specify the sampling space $\out$ anymore.

The POVM of a measuring apparatus gives the probability distribution of the outcomes for each input
state $\rho$ via the Born rule
\begin{equation}
p(i|\rho)\doteq\Tr[\rho P_i].
\end{equation}
The usual case of the customary observable corresponds to $\{P_i\}$ being the orthogonal projectors
on the eigen-spaces of a selfadjoint operator.
\par We now want to build up a detector which is "programmable", namely such that we can tune its
POVM by changing the state of an ancillary unit in the detector. Clearly, the most general
programmable detector would have its ancilla interacting with the measured system via a unitary
transformation $U$, which is then followed by an observable $\{E_i\}$ jointly measured on system and
ancilla, $U$ and $\{E_i\}$ being fixed constituents of the apparatus (due to the Naimark theorem,
considering a POVM in place of the observable $\{E_i\}$ would be simply equivalent to have a
higher dimensional ancilla and another fixed operator $U$). If such a detector could be
programmed to achieve a given POVM $\bP=\{P_i\}$ ideally,  this means that there would exist a 
"program state" $\sigma_\bP$ of the ancilla such that the following identity holds
\begin{equation}
p(i|\rho)=\Tr[\rho P_i]=\Tr[U(\rho\otimes\sigma_\bP)U^\dag E_i],\;\forall i,\forall\rho.
\label{eq:mathdescr}
\end{equation}
Clearly, the unitary interaction can be included in the definition itself of the joint observable
$\{E_i\}$ by defining $F_i\doteq U^\dag E_i U$ for 
all $i\in\out$. By taking the partial trace in Eq. \eqref{eq:mathdescr} over the ancilla and using
the polarization identity (Eq. \eqref{eq:mathdescr} holds for all states) one obtains
\begin{equation}
P_i=\Tr_A[(I\otimes\sigma_\bP)F_i].
\label{eq:prog}
\end{equation}
Therefore, a programmable detector is completely specified by the joint POVM $\bF=\{F_i\}$ on system
plus ancilla, therefore in the following the detector will be identified with $\bF$. Notice that
from Eq. (\ref{eq:prog}) it follows that the convex set of states $\anc$ of the ancilla is in correspondence 
via the map $\map{M}_\bF(\sigma)\doteq[(I\otimes\sigma)\bF]$ with a convex subset $\PF$ 
of the convex set $\Pn$ of the system POVM's with the same number $n\leqslant\infty$ of outcomes of $\bF$ 
($\PF$ is the convex set of POVM's that can be achieved with the fixed programmable detector $\bF$).
Therefore, if the POVM $\bP$ is extremal  (e.~g. it is an observable \cite{our}), and if there
exists a state of the ancilla $\sigma_\bP$ that satisfies identity (\ref{eq:mathdescr}), then there
also exists a pure state $\sigma_\bP$ satisfying the same identity: we will use this observation in the following.
\par The problem of measurement programmability can be restated in mathematical terms by asking if
$\PF\doteq\map{M}_\bF(\anc)\equiv\Pn$ for some $\bF$. In words: there exists a POVM
$\bF$ such that by varying the state $\sigma\in\anc$ in Eq. (\ref{eq:prog}) one obtains the full
convex set of POVM's $\Pn$ on $\sH$? We will show now that this is impossible, and we will use for
this purpose a generalization of the argument of Ref. \cite{fiurasek1}.  
\par Let us consider a two level system $\sH\simeq\Cmplx^2$, and suppose that
we want to program at least all possible observables by means a single programmable detector with 
finite-dimensional ancilla.  Each observable on $ \sH\simeq\Cmplx^2$ is simply a two-outcome
orthogonal POVM  $\{P,I-P\}$, with $P=|\psi\>\<\psi|$ and $I-P=|\psi^\perp\>\<\psi^\perp|$,
$|\psi\>$ being a unit vector in $\sH$. In other words, the observables on $ \sH\simeq\Cmplx^2$ 
are in correspondence with the pure states of the system. As previously noticed, without loss of
generality we can take the program state $\sigma_\psi$ as pure,  and we will denote
it as $\sigma_\psi=|\Phi(\psi)\>\<\Phi(\psi)|$. The POVM $\bF$ of the programmable detector would
then be a two-value POVM---so-called {\em effect}---$\{F,I-F\}$, and exact programmability for all
observables would imply  
\begin{equation}
|\psi\>\<\psi|=\Tr_A[(I\otimes |\Phi(\psi)\>\<\Phi(\psi)|)F],
\end{equation}
namely
\begin{equation}
\begin{split}
  & (\<\psi|\otimes\<\Phi(\psi)|)F(|\psi\>\otimes|\Phi(\psi)\>)=1\,,\\
  &(\<\psi|\otimes\<\Phi(\psi)|)F(|\psi^\perp\>\otimes|\Phi(\psi)\>)=\\
  &(\<\psi^\perp|\otimes\<\Phi(\psi)|)F(|\psi^\perp\>\otimes|\Phi(\psi)\>)=0\,.
\label{eq:condobqu}
\end{split}
\end{equation}
Equations \eqref{eq:condobqu} imply that for all $\psi\in\sH$ one has
\begin{equation}
F|\psi\>|\Phi(\psi)\>=|\psi\>|\Phi(\psi)\>,\qquad F|\psi^\perp\>|\Phi(\psi)\>=0,
\end{equation}
namely, for all $\psi\neq\psi^\prime\in\sH$ one has
\begin{equation}
\<\psi^\perp|\<\Phi(\psi)|F|\psi^\prime\>|\Phi(\psi^\prime)\>=\<\psi^\perp|\psi^\prime\>\<\Phi(\psi)|\Phi(\psi^\prime)\>=0,
\end{equation}
which implies that $\<\Phi(\psi)|\Phi(\psi^\prime)\>=0$. This
means that the ancillary system must have a continuum of orthonormal states, which cannot
happen in a separable Hilbert space. This proves that exact deterministic universal programmability of
measurements is impossible. 
\par One can then ask if it is possible to approximate all possible observables $\bP$ within some 
accuracy $\varepsilon^{-1}$ using a single finite-dimensional ancilla: here we will answer to this question
with a general lower bound for the optimal accuracy $\varepsilon^{-1}$ achievable by a programmable detector
versus the dimension of its ancilla. 
\par The first step is to give a precise definition of the accuracy of the approximation. For this
purpose, we consider the usual distance between two probability distributions $\{p_i\}$ and $\{q_i\}$
\begin{equation}
\delta({\bf p},{\bf q})=\sum_i|p_i-q_i|,
\end{equation}
and define accordingly the distance between two POVM's as the distance between their
respective probabilities, maximized over all possible states, namely
\begin{equation}
\delta({\bP},{\bQ})=\max_\rho\sum_i|\Tr[\rho(P_i-Q_i)]|\,.
\label{eq:dist}
\end{equation}
Then, we will say that the POVM $\bP$ approximates within $\varepsilon$ the POVM $\bQ$ if their distance
is less than $\varepsilon$, namely  
\begin{equation}
\delta({\bP},{\bQ})\leqslant\varepsilon.
\end{equation}
We will then rate the performance of a programmable detector $\bF$ saying that it achieves
accuracy $\varepsilon^{-1}$---shortly, it is {\em $\varepsilon$-programmable}---when
\begin{equation}
\max_{\bP\in\Pn}\min_{\bQ\in\PF}\delta({\bf P},{\bf Q})\leqslant\varepsilon.\label{minmax}
\end{equation}
\par We will now derive an upper bound for the function $d=f(\varepsilon)$ that gives the minimal
needed dimension of the ancilla to achieve accuracy $\varepsilon^{-1}$. We can restrict attention
to programmability of observables only, namely with $n=\dim(\sH)$ and $\Pn$ is substituted with the set of 
observables ${\cal O}_n$ in Eq. (\ref{minmax}). In fact, the generalization to nonorthogonal POVM's
is just equivalent to program observables in the larger dimension $n^2$ 
Clearly the function $d=f(\varepsilon^{-1})$ must be increasing, since the higher is the
accuracy $\varepsilon^{-1}$, the larger the minimal dimension $d$ needed for the ancilla, namely the
``size'' of the programmable detector.
\par The distance defined in Eq. \eqref{eq:dist} is hard to handle analytically, whence we bound it
as follows
\begin{equation}
\delta(\bP,\bQ)\leqslant\sum_i\N{P_i-Q_i}\leqslant\sum_i\N{P_i-Q_i}_2,
\label{eq:normbound}
\end{equation}
where $\N{A}$ is the usual operator norm of $A$, and $\N{A}_2\doteq\sqrt{\Tr[A^\dag A] }$ is the
Frobenius norm. Consider now a $d$-dimensional ancilla and a system-ancilla interaction $U$ of the
following {\em controlled-unitary} form
\begin{equation}
U=\sum_{k=1}^dW_k\otimes|\phi_k\>\<\phi_k|,\label{cntrW}
\end{equation}
where $\{\phi_k\}$ is an orthonormal complete set of vectors for the ancilla and $W_k$ are generic
unitary operators on $\sH$. Consider then a POVM $\bE=U\bF U^\dag$ of the form
\begin{equation}
E_i=|\psi_i\>\<\psi_i|\otimes I_A\,,
\end{equation}
where $I_A$ denotes the identity operator on the ancilla space, and $\{\psi_k\}$ is a complete
orthonormal set for the system. The observable to be approximated will then be written as follows
\begin{equation}
P_i=W^\dag|\psi_i\>\<\psi_i|W,\label{obsW}
\end{equation}
$W$ being a unitary operator on $\sH$, and we will scan all possible observables by varying $W$. 
For the  program state of the ancilla we use one of the states $\phi_k$, which give the POVM's 
\begin{equation}
Q_i=W_k^\dag|\psi_i\>\<\psi_i|W_k\,.
\label{eq:unitobs}
\end{equation}
This special form simplifies the calculation of the bound in Eq.  \eqref{eq:normbound}, which
becomes
\begin{equation}
\begin{split}
\delta(\bP,\bQ)&\leqslant\sum_i\sqrt{2(1-|\<\psi_i|W^\dag W_k|\psi_i\>|^2)}\\
&\leqslant\sqrt 2\sum_i\sqrt{2-\<\psi_i|(W^\dag W_k-W^\dag_k W)|\psi_i\>},
\end{split}
\end{equation}
and using the Jensen's inequality for the square root function we have
\begin{equation}
\delta(\bP,\bQ)\leqslant\sqrt{2n}\N{W-W_k}_2\,.
\label{eq:finbound}
\end{equation}
Now we can always take $d$ sufficiently large  such that we can choose the $d$ operators $\{W_k\}$
in the unitary transformation $U$ in Eq. (\ref{cntrW}) in such a way that for each given $W$ there
is always a unitary operator $W_k$ in the set for which $\sqrt{2n}\N{W-W_k}_2$ is bounded by
$\varepsilon$. This will guarantee that for the given observable $\bP$ corresponding to $W$ there is
a program state for the ancilla such that the POVM $\bQ$ achieved by the programmable detector is
close to the desired $\bP$ less than $\varepsilon$. The set of all possible unitary operators $W$
is a compact manifold of dimension $h=n^2-n$. We now consider a covering of the manifold with balls
of radius $r=\frac{\varepsilon}{\sqrt{2n}}$ centered at the operators $W_k$. This guarantees that any $W$
would be within a distance $\frac{\varepsilon}{\sqrt{2n}}$ from an operator $W_k$, which in turns
implies that the accuracy of the programmable device is bounded by $\varepsilon$ via
Eq. \eqref{eq:finbound}. Using the volume $V=\frac{\pi^{\frac{h}{2}}r^h}{\Gamma(\frac{1}{2}h+1)}$ of
the $h$-dimensional sphere of radius $r$, we obtain the number of balls needed for the covering (for
sufficiently small $\varepsilon$, corresponding to the upper bound for the minimal dimension of the
ancilla
\begin{equation}
d\leqslant\kappa(n)\left(\frac1\varepsilon\right)^{n(n-1)}\,,
\label{eq:polyn}
\end{equation}
where $\kappa(n)$ is a constant that depends on $n$. Eq. (\ref{eq:polyn}) gives an upper bound for
the dimension $d$ which is polynomial versus the accuracy $\varepsilon^{-1}$.
\par For qubits, the observable has only two elements, $P_0=|\psi\>\<\psi|$ and 
$P_1=|\psi_\perp\>\<\psi_\perp|=I-P_0$, and the distance in Eq. \eqref{eq:dist} can be evaluated
analytically as follows 
\begin{equation}
\delta(\bP,\bQ)=\max_\rho 2|\Tr[\rho(P_0-Q_0)]|\,.
\label{eq:distqub}
\end{equation}
The best device known\cite{fiurasek1} for programming qubit observables has
dimension of the ancilla which grows exponentially versus $\varepsilon^{-1}$. The programmable detector
uses $N$ qubits in the state $|\psi\>^{\otimes N}$, and the POVM $\bF=\{F_0,I-F_0\}$ is given by 
\begin{equation}
F_0=Z_+^{(N+1)},
\end{equation}
with $Z_+^{(N+1)}$ denoting the orthogonal projector over the totally symmetric Hilbert space
$(\sH^{\otimes N+1})_+$, where $\sH\simeq\Cmplx^2$. With this choice one can easily evaluate the
POVM programmed in the detector in Eq. (\ref{eq:prog}) obtaining
\begin{equation}
\begin{split}
Q_0=&\Tr_A[(I\otimes|\psi\>\<\psi|^{\otimes N})Z_+^{N+1}]\\
=&|\psi\>\<\psi|+\tfrac1{N+1}(I-|\psi\>\<\psi|)\,.
\end{split}
\end{equation}
Then, upon substituting $Q_0-P_0=\tfrac1{N+1}(I-|\psi\>\<\psi|)$ in Eq. \eqref{eq:distqub} one obtains
$\varepsilon\doteq\delta(\bP,\bQ)=\frac2{N+1}$, corresponding to
\begin{equation}
d=\tfrac12 4^{\varepsilon^{-1}},
\end{equation}
which must be compared with the polynomial growth in Eq. \eqref{eq:polyn}.
\par As regards now the programmability of all POVM's (i.~e. including the nonorthogonal ones),
just notice that one just needs to be able to program only the extremal POVM's in $\Pn$, since their
convex combinations will corresponds to mixing the program state or to randomly choosing among
different detectors. Then, since their maximum number of outcomes is $n^2$, the extremal POVM's have
Naimark's extension to observables in dimension $n^2$, whence we are reduced to the case of
programmability of observables in dimension $n^2$.
\par We will now give a programmable detector for qubits that achieves an accuracy that is linear in
$d$. For the ancilla we use a generic $d$-dimensional quantum system,
and relabel the dimension in the angular momentum fashion $d\doteq 2j+1$. The idea is now to design
a programmable detector in which the unitary transformation corresponding to the observable $\{P_i\}$
in Eq. (\ref{obsW}) is programmed by covariantly changing the program state of the ancilla. By labeling
unitary transformations by a group element $g\in\SU2$, we write the observable to be
programmed as $P_0\doteq V_g|\frac12\>\<\frac12|V_g^\dag$ where $\{V_g\}\equiv(\tfrac{1}{2})$ is a
unitary irreducible representation of $\SU2$ with angular momentum $\tfrac{1}{2}$, whereas the
program state will be written as  $W_g\sigma W_g^\dag$, with  $\{W_g\}\equiv(j)$ a unitary
irreducible representation of $\SU2$ on the ancilla space with angular momentum $j$. As
already noticed, without loss of generality we can always choose the state $\sigma$ as pure. We will
now show that a good choice for the program state is $\sigma=|j,j\>\<j,j|$, $\{|j,m\>\}$ denoting an
orthonormal basis of eigenstates of $J_z$ in the irreducible representation with angular momentum $j$. The
tensor representation $\{V_g\otimes W_g\}\equiv\tfrac{1}{2}\otimes j$
can be decomposed into the direct sum of two irreducible representations
$\tfrac{1}{2}\otimes j=j_+\oplus j_-$, where $j_\pm=j\pm \tfrac{1}{2}$. For the POVM $\bF$
of the programmable detector we will use $F_0=Z_+$ and $F_1=Z_-$, $Z_\pm$ denoting the orthogonal
projector on the invariant space for angular momentum $j_\pm$, i.~e.
\begin{equation}
F_0=\sum_{m=-j_+}^{j_+}\left|j_+, m\right\>\left\<j_+, m\right|.
\end{equation}
Using the invariance $(V_g\otimes W_g)F_0(V_g^\dag\otimes W_g^\dag)=F_0$, we can write
the programmed POVM as follows
\begin{equation}
\begin{split}
Q_0=&\Tr_A[(I\otimes W^\dag_g|j,j\>\<j,j|W_g)F_0]
\\=&V_g^\dag\Tr_A[(I\otimes|j,j\>\<j,j|)F_0]V_g
\\ =&V_g\left(\left|\tfrac12,\tfrac12\right\>\left\<\tfrac12,\tfrac12\right|
+\tfrac1{2j+1}\left|\tfrac12,-\tfrac12\right\>\left\<\tfrac12,-\tfrac12\right|\right)V_g^\dag,
\end{split}
\end{equation}
where we used the only non vanishing Clebsch-Gordan coefficients
$|\<j_+,j_+|\tfrac12,\tfrac12\>|j,j\>|^2=1$, and
$|\<j_+,j_-|\tfrac12,-\tfrac12\>|j,j\>|^2=\frac{1}{2j+1}$. Clearly,
$Q_0-P_0=\frac1{2j+1}V_g|\frac12,-\frac12\>\<\frac12,-\frac12|V_g^\dag$, whence according to
Eq. (\ref{eq:distqub}) the accuracy is given by $\delta(\bP,\bQ)=2/d$. The scaling of the dimension with the
accuracy is then linear
\begin{equation}
d=2\varepsilon^{-1},
\end{equation}
whereas the bound (\ref{eq:polyn}) would be quadratic $d\propto\varepsilon^{-2}$. 

\par We want to emphasize that the no-go theorem for programming observables  actually holds only
for universal programmability. Indeed, if, for example, we restrict programmability to
covariant POVM's, then exact deterministic programmability is possible. In fact, according to
Holevo theorem \cite{Holevo82} a general group-covariant POVM density has the form 
$P(\d g)=V_g\nu V_g^\dag\mu(\d g)$, with $\mu$ invariant measure on the group (for simplicity we
restrict to compact group and trivial stability group: a more general analysis can be found in
Refs. \cite{dar} and \cite{darchiri}). Then, it is easy to see that a necessary and sufficient condition in order to
have $P(\d g)$ positive and normalized is that the operator $\nu$ is positive and unit-trace, namely
a state. The POVM can then be programmed exactly using an ancilla with the same dimension of the
system and with program state $\transp{\nu}$, and using for the POVM $\bF$ the 
covariant Bell POVM $\{|V_g\kk\bb V_g|\}$ as one can easily check that $V_g\nu
V_g^\dag=\Tr_A[(I\otimes\transp{\nu})|V_g\kk\bb V_g|]$ 
[we used the notation $|V_g\kk\doteq\sum_{mn}(\<m|V_g|n\>)|m\>\otimes|n\>\in\sH^{\otimes2}$, and 
$\transp{\nu}$ as the transposed of $\nu$ with respect to the same basis used to define
$|V_g\kk$]. 
\par In conclusion, we have shown how it is possible to achieve deterministically a programmable
measurement with size polynomial in the accuracy. For qubits one can program observables with size linear 
in the accuracy, and for this we have provided and explicit example. Finally, we have noticed
that for covariant measurements exact programmability is feasible.  It is still an open problem what
is the actual minimal size of the programmable detector for given accuracy.
\par This work has been co-founded by the EC under the program ATESIT (Contract No.
IST-2000-29681) and the MIUR cofinanziamento 2003. P.P. acknowledges support from the INFM under
project PRA-2002-CLON. G.M.D. acknowledges partial support by the MURI program administered by
the U.S. Army Research Office under Grant No. DAAD19-00-1-0177.

\end{document}